%% file: arxiv.tex
\documentclass[journal,twocolumn]{IEEEtran}
\usepackage{fancyhdr}
\usepackage{amsmath,amssymb}
\usepackage{amsfonts}
\usepackage[dvips]{graphicx}
\usepackage{dsfont}
\usepackage{float}
\usepackage{algpseudocode}
\usepackage{algorithm}
\usepackage{comment}
\usepackage[hidelinks]{hyperref}    

\usepackage{color}
\usepackage{pgfplots}           
\usepgfplotslibrary{statistics}
\pgfplotsset{compat=1.17}
\usepackage{enumerate}
\usepackage{graphics}
\usepackage{cite}
\usepackage{mathrsfs}
\usepackage{stfloats}
\usepackage{xfrac}
\usepackage{tikz}
\usepackage{mathdots}
\usepackage{yhmath}
\usepackage{cancel}
\usepackage{color}
\usepackage{siunitx}
\usepackage{array}
\usepackage{multirow}
\usepackage{textcomp}
\usepackage{gensymb}
\usepackage{tabularx}
\usepackage{booktabs}
\usepackage{graphicx}
\usepackage{xcolor}

\newtheorem{lemma}{Lemma}

\begin{document}
	
	\title{Resolving the Double Near-Far Problem via Wireless Powered Pinching-Antenna Networks}
	
	\author{Vasilis K. Papanikolaou, Gui Zhou, Brikena Kaziu, Ata Khalili, Panagiotis D. Diamantoulakis, \\ George K. Karagiannidis, and Robert Schober,  
		\thanks{V. K. Papanikolaou, G. Zhou, B. Kaziu, A. Khalili, and R. Schober are with the Institute for Digital Communications (IDC), Friedrich-Alexander-University Erlangen-Nuremberg, Erlangen, Germany (emails: 
			\{vasilis.papanikolaou,gui.zhou,brikena.kaziu, ata.khalili,robert.schober\}@fau.de).}
		\thanks{P. D. Diamantoulakis and G. K. Karagiannidis are with the Wireless Communications \& Information Processing (WCIP) Group, Department of Electrical and Computer Engineering, Aristotle University of Thessaloniki, Thessaloniki, Greece (e-mails: \{padiaman,geokarag\}@auth.gr).}
		\vspace{-8mm}
	}
	\maketitle
	
\begin{abstract}
	This letter introduces a novel wireless powered communication system, referred to as a wireless powered pinching-antenna network (WPPAN), utilizing a single waveguide with pinching antennas to address the double near-far problem inherent in wireless powered networks. In the proposed WPPAN, users harvest energy from spatially distributed pinching antennas in the downlink and use the collected power to transmit messages in the uplink. Furthermore, to manage the combinatorial complexity associated with activating the pinching antennas, we propose three approaches of varying complexity to simplify the original resource allocation problem and then solve it efficiently using convex optimization methods. Simulation results confirm that the proposed WPPAN system effectively mitigates the double near-far problem by providing antenna resources closer to the users, thereby enhancing both downlink energy harvesting and uplink data transmission.
\end{abstract}
\vspace{-0.3cm}
\section{Introduction}
The rapid advancement of wireless communications and the need to serve a massive number of energy-constrained mobile devices has propelled the notion of energy harvesting to the forefront of technological advancement, aiming to reduce the need for frequent battery maintenance. In this context, wireless energy transfer from controllable RF sources has been proposed to ensure continuous device operation \cite{psomas}.  Wireless powered networks (WPNs) have emerged as a promising paradigm for enabling energy-constrained wireless devices to operate sustainably by using the harvest-then-transmit protocol. However, one of the critical challenges in these networks is the double near-far problem, where users located farther away from the energy source not only harvest less power but also experience harsher uplink channel conditions, thereby exacerbating performance disparities among users \cite{vranas, ruizhang,elenaMIMO}.

To this end, recent innovations in flexible-antenna systems, such as reconfigurable intelligent surfaces (RISs), fluid antennas, and movable antennas offer the ability to dynamically adjust the wireless channel to improve performance \cite{yu,fluid}. RIS systems can effectively mitigate line-of-sight (LoS) blockage but on the other hand, they introduce double attenuation. Fluid antennas, made of liquid metals, and movable antennas can change their positions and radiation patterns, but such changes are usually insufficient to mitigate significant path loss, especially for high carrier frequencies. 

Pinching-antenna systems are a novel class of flexible-antenna technology that use dielectric waveguides and small, reconfigurable dielectric elements to dynamically create and adjust radiation sites, thereby making wireless channels highly adaptable \cite{ding2024flexible}. Pinching elements are positioned accordingly along the waveguide and offer the ability to establish robust LoS links close to the served user \cite{tegos2025pinching, zhou2025} or sensing target \cite{bozanis2025,khalili2025}. Introduced by NTT DoCoMo in 2022 \cite{docomo}, pinching-antenna systems are therefore promising for alleviating the double near-far problem in WPNs.  

In this letter, we propose a novel wireless powered pinching-antenna network (WPPAN) designed to mitigate the double near-far problem. In the considered system, users first harvest energy from distributed pinching antennas and subsequently utilize the harvested energy to transmit data in the uplink. Time-division multiple access (TDMA) is adopted to enable the communication of multiple users.
A key challenge in the considered network arises from the combinatorial complexity associated with the activation of pinching antennas. To address this issue, we employ three  approaches of varying complexity that circumvent the intractability of the original problem by transforming it into a series of convex optimization problems, taking into account the particularities of pinching-antenna networks. 

To ensure a fair comparison with conventional systems, we benchmark our design against a multiple-input single-output (MISO) system that employs a multi-antenna transmitter having the same total number of antennas as the considered pinching-antenna system. This direct comparison allows us to isolate the benefits stemming from the unique spatial distribution and proximity of the pinching antennas to the users.
Our results demonstrate that the pinching-antenna system effectively overcomes the double near-far problem by providing antenna resources close to the users, thereby enhancing both downlink energy harvesting and uplink data transmission.

\vspace{-0.5cm}
\section{System Model}
\subsection{Pinching-Antenna System}
We consider a wireless network consisting of a base station (BS) equipped with a single waveguide with $N$ pinching antennas and $M$ single-antenna devices, whose indices are collected in sets $\mathcal{N} = \{1, \dots, N\}$ and $\mathcal{M} = \{1, \dots, M\}$, respectively.
The $N$ pinching antennas are placed at fixed positions on the waveguide but their activation can be controlled. Considering a three-dimensional coordinate system, we assume that the devices are randomly deployed in a rectangular area lying in the $x$-$y$ plane with side-lengths $D_x$ and $D_y$, where $\boldsymbol{\psi}_m = (x_m, y_m, 0)$ denotes the position of the $m$-th device. Specifically, $x_m$ follows a uniform distribution in $[-D_x/2, D_x/2]$ and $y_m$ follows a uniform distribution in $[-D_y/2, D_y/2]$, respectively. 

For the pinching-antenna system, it is assumed, that the waveguide is installed at height $d$. The position of the $n$-th pinching antenna is $\boldsymbol{\psi}_P^n = (x_P^n, y_P^n, d)$, where $x_P^n \in [-D_x/2, D_x/2]$ and $y_P^n = 0,\,\forall n\in\mathcal{N}$. The LoS channel between the $m$-th device and the $N$ pinching antennas is given by
\begin{equation}
	\mathbf{h}_{m,1} =
	\left[
	\frac{{\eta} e^{-j \frac{2\pi}{\lambda} |\boldsymbol{\psi}_m - \boldsymbol{\psi}_P^1|}}{|\boldsymbol{\psi}_m - \boldsymbol{\psi}_P^1|}, \dots, 
	\frac{\eta e^{-j \frac{2\pi}{\lambda} |\boldsymbol{\psi}_m - \boldsymbol{\psi}_P^N|}}{|\boldsymbol{\psi}_m - \boldsymbol{\psi}_P^N|}
	\right]^\mathrm{T},
	\label{eq:channel_device}
\end{equation}
where $\eta = \frac{c}{4\pi f_c}$ is a constant with $c$, $f_c$, and $\lambda$ denoting the speed of light, the carrier frequency, and the wavelength in free space, respectively, and $(\cdot)^\mathrm{T}$ denotes the matrix transpose. Since all $N$ pinching antennas are positioned along the same waveguide, the signal transmitted by any antenna is essentially an attenuated phase-shifted replica of the signal transmitted by the BS at the feed point of the waveguide. Thus, the channel in the waveguide is given by
\begin{align}
	\mathbf{h}_2 = 
	\left[ 
	10^{-\frac{\kappa |\boldsymbol{\psi}_P^0 - \boldsymbol{\psi}_P^1|}{20}} e^{-j \frac{2\pi}{\lambda_g} |\boldsymbol{\psi}_P^0 - \boldsymbol{\psi}_P^1|}, \dots,\right. \nonumber \\ \left. 
	10^{-\frac{\kappa |\boldsymbol{\psi}_P^0 - \boldsymbol{\psi}_P^N|}{20}}
	e^{-j \frac{2\pi}{\lambda_g} |\boldsymbol{\psi}_P^0 - \boldsymbol{\psi}_P^N|}
	\right]^\mathrm{T},
	\label{eq:channel_waveguide}
\end{align}
where  $\boldsymbol{\psi}_P^0$ denotes the position of the feed point of the waveguide, $\kappa$ is the average attenuation factor along the dielectric waveguide in dB/m \cite{tegos2025pinching}, and $\lambda_g = \lambda / n_\text{eff}$ denotes the guided wavelength with $n_\text{eff}$ being the effective refractive index of the dielectric waveguide. Then, the total channel between the $m$-th device and the $N$ pinching antennas is given as $\mathbf{h}_m = \mathbf{h}_{m,1}\odot\mathbf{h}_2$, where $\odot$ denotes the Hadamard product.

\subsection{Network Protocol}
We assume the communication is divided into time frames of duration $T$ and the channel state remains constant during a time frame and can be perfectly estimated by the BS. We further assume that the network adopts a harvest-then-transmit protocol, i.e., $\tau_0,\, 0\leq \tau_0\leq T$, amount of time is assigned to the BS to broadcast wireless energy to all users, while the remaining time, $T-\tau_0$, is available for data transmission by the users. The transmission of the users is limited by the total energy they have harvested during the first phase. We refer to the proposed system as WPPAN.

\subsection{Non-Linear Energy Harvesting Model}
Since the linear energy harvesting model fails to capture the non-linearities of the underlying hardware, we adopt the non-linear model from \cite{elenaMIMO}. Therefore, the $m$-th user's total harvested energy $\Phi_{\mathrm{EH},m}$ is given by
\begin{equation}
	\label{eq: harvestingmodel}
	\Phi_{\mathrm{EH},m} = P_\mathrm{EH,max} \frac{1 - e^{-a_\mathrm{EH} P_{\mathrm{EH},m}}}{1 + e^{-a_\mathrm{EH}(P_{\mathrm{EH},m} - b_\mathrm{EH})}}, \quad m\in\mathcal{M},
\end{equation}
where $P_{\mathrm{EH,max}}$ denotes the maximum power that the receiver can harvest, and $a_\mathrm{EH}$ and $b_\mathrm{EH}$ model physical hardware properties \cite{elenaMIMO}.
Furthermore, by denoting the effective channel gain of the $m$-th user as $g_{m,i}$, the total received RF power of the $m$-th user is given by $P_{\mathrm{EH}, m} =  |g_{m}|^2P_0,$ where $P_0$ is the transmit power of the BS. Here, $g_{m}$ includes the channel effects in \eqref{eq:channel_device} and \eqref{eq:channel_waveguide}. In the next section, the effective channel gain will be presented in more detail for the considered WPPAN.

\section{Wireless Powered Pinching-Antenna Networks}
In this section, we examine the particularities of WPPANs and present a corresponding harvest-then-transmit protocol. 
\subsection{Antenna Activation}
First, we define the set $\mathcal{Q}$ that contains all possible subsets $\mathcal{K}_q$ of $\mathcal{N}$ that can be utilized in the downlink phase. $\mathcal{Q}$ is directly related to the possible pinching-antenna combinations that can be used, i.e., each subset $\mathcal{K}_q$ denotes one possible combination. A timeslot duration $\tau^d_q$ within $\tau_0$ is assigned to each subset $\mathcal{K}_q\in\mathcal{Q}$ so that $\sum_{q=1}^{|\mathcal{Q}|} \tau_q^d \leq \tau_0$, where $|\mathcal{Q}|$ denotes the cardinality of set $\mathcal{Q}$. If $\tau_q^d=0$ for some $q$, that means the particular pinching-antenna combination $\mathcal{K}_q$ is not utilized during the entire frame.
During the second phase of the protocol proposed in Section II.B, the uplink transmission from the users to the BS takes place. In this work, we assume TDMA is utilized and each user $m$ is allocated a timeslot $\tau^u_m$ so that $\sum_{m=1}^M \tau^u_m \leq T - \tau_0$. 

In each uplink timeslot, $\tau_m, \,m\in\mathcal{M}$, a different subset of pinching antennas can be active. Fundamentally, we assume that the channel is reciprocal, so the channel gain between a specific pinching antenna and a specific user is the same in downlink and uplink. The effective channel gains of a user in the first phase and the second phase of transmission are generally not identical because the effective channel gain is dependent on which pinching antennas are active in each timeslot. Additionally, in the downlink, the power is equally distributed among the active pinching antennas, but in the uplink, antenna noise is accumulated over the active pinching antennas. Assuming antenna noise, not the noise from the RF chain, is the main source of impairment for a given subset of activated antennas, the signal-to-noise-ratios (SNRs) in uplink and downlink are identical \cite{tegos2025pinching}. Therefore, channel reciprocity holds for the effective channel gains in uplink and downlink, provided the same pinching antennas are active. Thus, in timeslot $\tau_\zeta^\ell$, for user $m$, the effective channel gain
\begin{equation}\label{eq:channel_downlink}
	g_{m,\zeta}^\ell = \frac{\sum_{n=1}^N b^\ell_{\zeta,n} h_{m,1,n} h_{2,n}}{\sqrt{\sum_{n=1}^Nb^\ell_{\zeta,n}}},
\end{equation}
where $\ell\in\{d,u\}$, $\zeta = q\mathds{1}_{\ell=d}+m\mathds{1}_{\ell=u}$ with $\mathds{1}_{\{\cdot\}}$ being the indicator function, $b^\ell_{\zeta,n}\in\{0,1\}$ denotes the activation indicator of pinching antenna $n$ in timeslot $\tau_\zeta^\ell$, with at least one antenna active so the denominator in \eqref{eq:channel_downlink} is never $0$, and $h_{m,1,n}$ and $h_{2,n}$ are the $n$-th elements of $\mathbf{h}_{m,1}$ and $\mathbf{h}_2$, respectively. 

\subsection{Energy Harvesting}
Based on the effective channel gain definition in \eqref{eq:channel_downlink}, in this subsection, we revisit the energy harvesting. The total received RF power at user $m$ during downlink timeslot $\tau_q^d$ can be expressed as
\begin{equation}
	P_{\mathrm{EH}, m,q} = |g_{m,q}^d|^2 P_0 = \frac{|\mathbf{h}_m^\text{T}\mathbf{b}_q^d|^2}{\mathbf{1}^\text{T}\mathbf{b}_q^d}P_0,
\end{equation}
where $\mathbf{b}^d_q=\{b^d_{q,n}\}_{n=1}^{N}$ is the downlink activation vector in timeslot $q$.

The energy harvested during $\tau_0$ by user $m$ is given as
\begin{equation}\label{eq:harvestedenergy}
	E_m = \sum_{q=1}^{|\mathcal{Q}|} \tau^d_q \Phi_{\mathrm{EH},m,q}(\mathbf{b}_q^d),
\end{equation}
where $\Phi_{\mathrm{EH},m,q}$ is calculated based on \eqref{eq: harvestingmodel} and the index $q$ is added to indicate that the effective channel gain is $g_{m,q}^d$.

\subsection{Information Transmission}
Since each user aims to utilize all of their harvested energy for information transmission, given \eqref{eq:harvestedenergy}, the power consumed for communication is constrained by $P_m = E_m/\tau^u_m$. The data rate of user $m$ in uplink timeslot $\tau_m^u$ is then given by
\begin{align}
	R_m &= \tau^u_{m} \log_2\left(1 +
	\frac{\rho|g_{m}^u|^2}{\tau^u_{m}}\sum_{q=1}^{|\mathcal{Q}|}\tau_{q}^d \Phi_{\mathrm{EH},m,q}\right),
\end{align}
where $\rho = 1/\sigma^2$, with $\sigma^2$ being the variance of the additive white Gaussian noise (AWGN).

\section{Minimum Rate Maximization}
We focus on maximizing the minimum data rate, which ensures user fairness. To do so, we optimize the activation vectors, $\mathbf{b}_q^d,\,\forall q\in\{1,\dots,|\mathcal{Q}|\}$ and $\mathbf{b}^u_m\,\forall m\in\mathcal{M}$, of the WPPAN, and the timeslot durations in the first and second phases, i.e., $\boldsymbol{\tau}_d\triangleq \{\tau_q^d\}_{q=1}^{|\mathcal{Q}|}$ and $\boldsymbol{\tau}_u\triangleq \{\tau_m^u\}_{m=1}^{M}$. The resulting optimization problem can be expressed as follows

\begin{equation}
	\begin{array}{ll}
		\underset{\mathbf{b}^d_q, \mathbf{b}_m^u,\boldsymbol{\tau}_d,\boldsymbol{\tau}_u}{\mathrm{max}}& \underset{m\in\mathcal{M}}{\mathrm{{min}}} R_m \\
		\quad\quad\!\mathrm{s.t.}& \mathrm{C}_1: \sum_{q=1}^{|\mathcal{Q}|} \tau_q^d + \sum_{m=1}^M \tau_m^u \leq T,\\
		& \mathrm{C}_2: \{\tau_q^d, \tau_m^u\} \geq 0, \,\forall m\in\mathcal{M}, \\ & \qquad \forall q\in\{1,\dots,|\mathcal{Q}|\}.
	\end{array}
	\label{opt TS}
\end{equation}

Problem \eqref{opt TS} is non-convex due to the discrete variables $\mathbf{b}_q^d$ and $\mathbf{b}_m^u$, the use of the $\min$ function, as well as the coupling of $\boldsymbol{\tau}_d$ and $\boldsymbol{\tau}_u$ with the activation vectors $\mathbf{b}_q^d$ and $\mathbf{b}_m^u$. To overcome these challenges, first, we transform the optimization problem in \eqref{opt TS} in its equivalent epigraph form with $v$ being the hypograph variable. This leads to
\vspace{-0.2cm}
\begin{equation}
	\begin{array}{ll}
		\underset{\mathbf{b}^d_q, \mathbf{b}_m^u,\boldsymbol{\tau}_d,\boldsymbol{\tau}_u,v}{\mathrm{{max}}}&  v \\
		\quad \quad\mathrm{{s.t.}}& \mathrm{C}_1, \, \mathrm{C}_2\\
		& \mathrm{C}_3:  v\leq  \tau^u_{m} \log_2\left(1 +
		\frac{\rho|g_{m}^u(\mathbf{b}_m^u)|^2}{\tau^u_{m}} \right. \\
		& \quad\quad \left.\times \sum_{q=1}^{|\mathcal{Q}|}\tau_{q}^d \Phi_{\mathrm{EH},m,q}(\mathbf{b}_q^d)\right), \,\forall m\in\mathcal{M}, \\ & \qquad \forall q\in\{1,\dots,|\mathcal{Q}|\}.
	\end{array}
	\label{opt TS3}
\end{equation}
Problem \eqref{opt TS3} is still a mixed integer non-linear problem (MINLP), and as $\mathbf{b}_q^d$ and $\mathbf{b}_m^u$ are discrete, it has a combinatorial nature, making it NP-hard. In the following, we propose three approaches of varying complexity to efficiently decouple the optimization variables and maximize the minimum rate in the WPPAN.

\subsection{Search-based WPPAN}
In this approach, which we name search-based WPPAN, we set $\mathcal{Q} = \mathcal{S} \triangleq 2^\mathcal{N}$, i.e., the power set containing all subsets of $\mathcal{N}$. Thus, all possible combinations of antenna activations are considered for the downlink. Therefore, $\mathbf{b}_q^d$ can be omitted from the optimization variables in \eqref{opt TS3}. Additionally, in the uplink, each user $m$ selects the pinching-antenna activation vector $\mathbf{b}_m^u$ that maximizes their SNR given as $\gamma_m = \frac{\rho|g_{m}^u(\mathbf{b_m^u})|^2}{\tau^u_{m}}\sum_{q=1}^{|\mathcal{S}|}\tau_{q}^d \Phi_{\mathrm{EH},m,q}$. Search‑based WPPAN attains the optimal solution to \eqref{opt TS} because the set of downlink activation vectors 
$\mathcal B^{d}=\{\mathbf b^{d}_{q}\}_{q=1}^{|\mathcal{Q}|}$ exhaustively enumerates every antenna subset in $\mathcal S$, and, in the uplink, each user $m$ chooses the activation vector $\mathbf b^{u}_{m}$ that maximises its individual data rate. The latter can be understood by examining $
\mathrm{C}_3$ of \eqref{opt TS3}. Since $\tau_m^u$ and $\gamma_m$ are positive, due to the monotonicity of the logarithm, and by noticing that $\gamma_m$ is not affected  by $\mathbf{b}_\mu^u,\,\mu\neq m$, maximizing $|g_{m}^u(\mathbf{b}_m^u)|$ does not affect any other constraint or variable.
Thus, the optimal antenna activation vector for uplink timeslot $m$, i.e., when the $m$-th user is transmitting, is given by
\begin{equation} \label{eq:optimaluplink}
	{\mathbf{b}_m^u}^* = \underset{\substack{\mathbf{b}_m^u\in \{0,1\}^{N\times 1}}}{\mathrm{argmax}} \quad |g_{m}^u(\mathbf{b}_m^u)|^2.
\end{equation}
Problem \eqref{opt TS3} can then be solved optimally, as described in the following Lemma.
\begin{lemma}\label{lemma:1}
For fixed $\mathbf{b}_u$, given by \eqref{eq:optimaluplink}, and $\mathcal{Q} = \mathcal{S}$, \eqref{opt TS3} is reduced to a convex optimization problem in $\boldsymbol{\tau}_d$ and $\boldsymbol{\tau}_u$ and can therefore be solved optimally in polynomial time with interior-point methods.
\end{lemma}
\begin{IEEEproof}
The proof is provided in Appendix \ref{app:1}.
\end{IEEEproof}
It should be noted that the size of $\boldsymbol{\tau}_d$ in \eqref{opt TS3} grows exponentially with $N$ and even though it is generally understood that convex problems can be solved efficiently in polynomial time, in this case, it can become challenging. 
\vspace{-0.2cm}

\subsection{Greedy WPPAN}
Since the number of variables in the first phase grows exponentially with the number of pinching antennas, a suboptimal but more efficient way to solve this problem is to limit the number of timeslots in the first phase to $M$. Furthermore, the antenna activation for $\tau_m^d$ is selected as the one maximizing the effective downlink channel gain of user $m$, but energy is broadcasted again to all users during the first phase. Then, the optimal selection for both uplink and downlink is obtained from \eqref{eq:optimaluplink} as ${\mathbf{b}_{m}^{d}} ={\mathbf{b}_{m}^{u}} = {\mathbf{b}_m^u}^* $, for all $m\in \mathcal{M}$.
Therefore, $\mathbf{b}_q^d$ and $\mathbf{b}_m^u$ are set and the rest of the greedy WPPAN protocol follows problem \eqref{opt TS3}. More specifically, this can be solved in the same way as the search-based WPPAN but by replacing $|\mathcal{Q}|$ by $M$.
\subsection{Naive WPPAN}
The greedy WPPAN method partially avoids the complexity of the search-based WPPAN by significantly reducing the number of optimization variables. However, solving \eqref{eq:optimaluplink} still requires searching through vectors whose sizes grow exponentially with the number of pinching antennas. We propose the naive WPPAN protocol, based on a simplified greedy WPPAN to eliminate the remaining combinatorial search by forcing every slot to activate \emph{exactly one} pinching antenna, i.e., $\mathbf 1^{\mathsf T}\mathbf b^{d}_{m}=\mathbf 1^{\mathsf T}\mathbf b^{u}_{m}=1$. With this one‑hot constraint the activation vectors $\mathbf b^{d}_{m}$ and  $\mathbf b^{u}_{m}$ are found by a simple $N$‑element scan. Finally, with $\mathbf{b}_m^d$ and $\mathbf{b}_m^u$ set, the rest of the naive WPPAN method follows problem \eqref{opt TS3} by setting the rest of the unused timeslot durations to zero. With only the timeslot duration vectors $\boldsymbol{\tau}_d$ and $\boldsymbol{\tau}_u$ being optimization variables, problem \eqref{opt TS3} becomes convex.

The three proposed methods to optimize the WPPAN are summarized in Algorithm \ref{alg:WPPAN}.
\begin{algorithm}
\caption{WPPAN Resource Allocation Engine}\label{alg:WPPAN}
{\small \begin{algorithmic}[1]
	\State \textbf{Input:}  $\mathcal N$, $\mathcal M$, $T$, $P_0$, channel vectors $\mathbf h_{m,1}$ and $\mathbf h_2$
	\State \textbf{Choose method} $\mathsf{Mode}\in\{\textsc{Search},\textsc{Greedy},\textsc{Naive}\}$
	\If{$\mathsf{Mode}=\textsc{Search}$}   \Comment{full power set}
	\State $\mathcal Q \gets 2^{\mathcal N}$, \quad $\mathcal B^{d}\gets\{\mathbf b^{d}_q\}_{q=1}^{|Q|}$.  \Comment{enumerate all}
	\State $\mathbf b^{u}_m\leftarrow$  as in \eqref{eq:optimaluplink}, $\forall m\in\mathcal{M}.$
	\ElsIf{$\mathsf{Mode}=\textsc{Greedy}$}   
	\State $\mathcal Q\leftarrow\mathcal M$. \Comment{one DL slot per user}
	\State $\mathbf b^{u}_m\leftarrow$ as in \eqref{eq:optimaluplink}, $\forall m\in\mathcal{M}.$
	\State $\mathbf b^{d}_m\leftarrow \mathbf b^{u}_m$, $\forall m\in\mathcal{M}$.
	\ElsIf{$\mathsf{Mode}=\textsc{Naive}$}   
	\State $\mathbf b^{u}_m\leftarrow$  one‑hot that maximizes $|\mathbf h_{m,1}^{\mathsf T}\mathbf b|$.
	\State $\mathcal Q\leftarrow\mathcal M$.
	\State $\mathbf b^{d}_m\leftarrow \mathbf b^{u}_m$. 
	
	\EndIf
	\State Assemble variables $\boldsymbol{\tau}=\{\tau^{d}_{q},\tau^{u}_{m}\}$ and solve \eqref{opt TS3} with a convex solver, e.g., CVX.
	\State \textbf{Output:} Optimized $\boldsymbol{\tau}^{d}_{q},\boldsymbol{\tau}^{u}_{m}$ and $\mathbf b^{d}_{q},\mathbf b^{u}_{m}$.
	\end{algorithmic}}
\end{algorithm}

\vspace{-0.7cm}
\section{Numerical Results}
In this section, the performance of the WPPAN with the proposed method is evaluated. 
We define the following simulation scenario. In a space defined by $(x,y)\in[-D_x/2,D_x/2]\times[-D_y/2,D_y/2]$, we place a waveguide at height $d$ on the line segment $y=0$ from $x_\mathrm{start} = -D_x/2$ to $x_\mathrm{end} = D_x/2$ with $N$ uniformly placed pinching antennas, which is connected to a BS located at $(x_\mathrm{start},0,d)$.
The system parameters, unless specified otherwise, are given as follows. The number of users is \(M = 3\), uniformly distributed in the room, and the number of antennas is \(N = 4\). The transmit power is \(P_0 = 40\) dBm, while the noise power is \(\sigma^2 = -95\) dBm. The room dimensions are $D_x=10$ m, $D_y=10$ m and the height is $d=3$ m. Rician fading with Rician factor $K_\text{Rician}=10$ is assumed throughout. The energy harvesting parameters are \(a_{\rm EH} = 1500\) and \(b_{\rm EH} = 0.0022\). The parameters are consistent with \cite{ding2024flexible} and \cite{mmWaveEH}, where the authors studied a mmWave WPN. The carrier frequency is \(f_c = 28\) GHz, and the refractive index in the waveguide is \(n_\text{eff} = 1.4\). 1000 Monte Carlo trials are conducted. 
We compare the proposed pinching-antenna system to a conventional fixed-antenna WPN where connected to the BS are $N$ transmit antennas. The reference WPN supports full digital beamforming \cite{elenaMIMO}.

\begin{figure}
\centering
\input{vsP0}
\caption{Maximum minimum rate versus transmit power.}\label{fig:p0}
\end{figure}
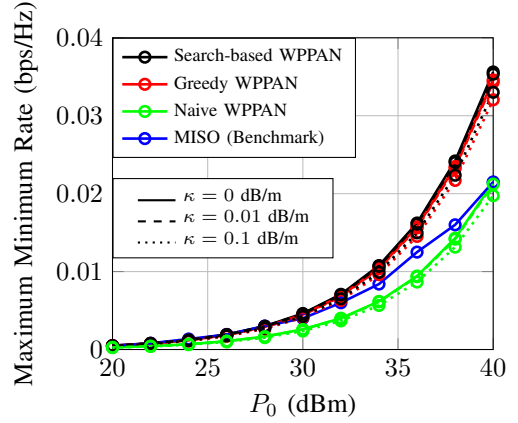

In Fig. \ref{fig:p0}, the maximum minimum rate is shown as a function of the transmit power. For low transmit powers $P_0$ all schemes yield data rates that are nearly zero; as $P_0$ rises, however, the proposed WPPAN shows clear superiority, ultimately delivering approximately 50 \% higher rates than the benchmark MISO WPN.
The greedy WPPAN appears to be near-optimal as it is very close to the search-based solution, indicating that only a few combinations of active antennas are effectively used. By contrast, the low-complexity naive WPPAN falls behind the other variants, yet at high $P_0$ it shadows the MISO benchmark almost exactly while demanding far less computation, showing that the full WPPAN gain appears only when \emph{pinching beamforming} is employed by carefully activating multiple pinching antennas. Additionally, the effect of waveguide losses is also shown. Despite the losses in the waveguide, for standard values \cite{ding2024flexible}, the performance is only slightly reduced and search-based and greedy WPPANs still surpass the MISO benchmark by a large margin.

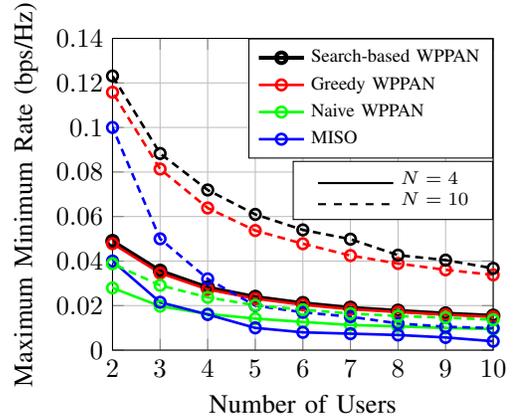
\begin{figure}
\centering
\input{vsUsers}
\caption{Maximum minimum rate versus number of users.}\label{fig:users}
\end{figure}

In Fig. \ref{fig:users}, the maximum minimum rate is investigated as a function of the number of users in the system. As expected, the performance of the system drops as more users join, as we maximize the rate for the worst-case user. By adding more pinching antennas, their placement becomes more dense, which leads to improved performance. Interestingly, for the naive WPPAN, adding more pinching antennas does not offer as large performance gains as for the search-based and greedy WPPANs, indicating again the importance of pinching beamforming. Even with larger number of pinching antennas, the greedy WPPAN appears to be near-optimal. Additionally, the the benchmark MISO system, due to the double near-far effect, cannot match the performance of the pinching-antenna system, especially when more users join the system.

\begin{figure}
\centering
\input{histogram}
\caption{Sampled probability density distribution of optimal number active pinching antennas for $N=10$, $M=5$.}\label{fig:histogram}
\end{figure}
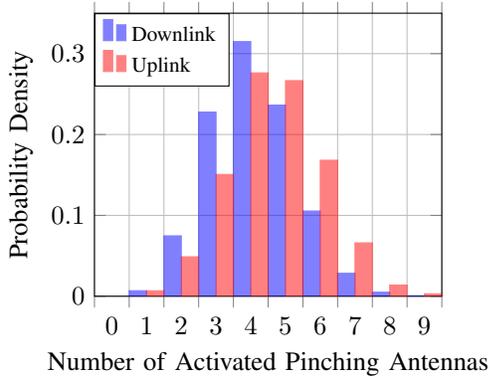
Fig. \ref{fig:histogram} reveals that on average fewer pinching antennas are activated in the downlink than in the uplink. This can be explained by the fact that all $M$ devices can harvest energy during all downlink time slots, whereas during each uplink timeslot, only one user can be served. Therefore, while highly selective beamforming, facilitated by a large number of activated antennas, is beneficial in the uplink, this is not necessarily true downlink, where each transmission benefits all users.

\vspace{-0.6cm}
\section{Conclusions}
A flexible-antenna system based on pinching antennas was investigated for use in WPNs. The proposed system, termed WPPAN, offers flexibility by distributing pinching antennas in space. This was shown to effectively alleviate the performance drop due to the double near-far problem in WPNs. While the combinatorial nature of the optimal solution in search-based WPPAN makes the problem intractable for networks with many pinching antennas, the greedy scheme remains near-optimal.
\vspace{-0.6cm}

\appendices
\section{Proof of Lemma \ref{lemma:1}}\label{app:1}
The objective function, constraints $\mathrm{C}_1$ and $\mathrm{C}_2$ are linear functions. Constraint $\mathrm{C}_3$ is a function of $\tau_m^u$ and $\tau_q^d, \, \forall q\in\{1,\dots,|\mathcal{Q}|\}$ and a linear function of $v$. For ease of notation, we rewrite it as 
\begin{equation}
w =  \tau_m^u \log_{2}\left(1 + 
\frac{\Psi}{\tau_m^u}\sum_{q=1}^{|\mathcal{Q}|}\tau_q^d\Phi_{\mathrm{EH},m,q}\right) - v,
\end{equation}
where $\Psi = \rho |g_m^u|^2$. We are interested in showing that $w$ is concave in $\boldsymbol{\tau}_d$ and $\boldsymbol{\tau}_u$. To help with the proof, we have the affine mapping $\varpi = \sum_{q=1}^{|\mathcal{Q}|}\tau_q^d\Phi_{\mathrm{EH},m,q}$. We consider the following function $\hat{w}$ 
\begin{equation}
\hat{w} = \tau_m^u \log_2\left(1 +
\frac{\varpi}{\tau_m^u}\Psi\right).
\end{equation}
By taking the Hessian matrix $\nabla^2{\hat{w}}$ of $\hat{w}$ we get
\begin{equation}
\nabla^2\hat{w}(\varpi, \tau_m^u) = \begin{bmatrix}
\frac{\partial^2 \hat{w}}{\partial\varpi^2} & \frac{\partial^2 \hat{w}}{\partial\varpi\partial\tau_m^u} \\
\frac{\partial^2 \hat{w}}{\partial\varpi\partial\tau_m^u} & \frac{\partial^2 \hat{w}}{\partial{\tau_m^u}^2}
\end{bmatrix},
\end{equation}
which after a few algebraic manipulations can be shown to be equal to
\begin{equation}
\nabla^2\hat{w}(\varpi, \tau_m^u) = \frac{\Psi^4}{(\Psi^2\varpi + \tau_m^u)^2\log(2)}\begin{bmatrix}
-\tau_m^u &  \varpi \\
\varpi & -\frac{\varpi^2}{\tau_m^u}
\end{bmatrix},
\end{equation}

where $\log(\cdot)$ denotes the natural logarithm. The eigenvalues of $\nabla^2\hat{w}(\varpi, \tau_m^u)$ can be calculated easily as
\begin{equation}
\lambda_{\hat{w},1} = -\frac{\Psi^4(\varpi^2 + {\tau_m^u}^2)}{\tau_m^u(\Psi^2\varpi + \tau_m^u)^2\log(2)}
\end{equation}
and $\lambda_{\hat{w},2} = 0.$
Since $\lambda_{\hat{w},1},\, \lambda_{\hat{w},2} \leq 0$ for the regions where $\boldsymbol{\tau}_d$ and $\boldsymbol{\tau}_u$ are defined, the Hessian matrix is proven to be negative semi-definite. Therefore, function $\hat{w}$ is concave. Since $\varpi=\sum_{q=1}^{|\mathcal{Q}|}\tau_q^d\,\Phi_{\mathrm{EH},m,q}$ is an affine mapping of the positive variables $\tau_q^d,\,\forall q\in\{1,\dots,|S|\}$, the monotonicity of $\hat{w}$ in $\varpi$ ensures that the overall composition $\hat{w}(\varpi(\{\tau_q^d\}),\tau_m^u)$ is concave in all of its arguments. Therefore, $\hat{w}$ is concave on its domain and the proof is completed.
\vspace{-0.5cm}
\bibliographystyle{IEEEtran}
\bibliography{refs.bib}

\end{document}

%% file: vsP0.tex
\begin{tikzpicture}
        \begin{axis}[
        clip = false,
            scaled ticks=false, 
            tick label style={/pgf/number format/fixed},
            width=0.75\linewidth,
            xlabel = {$P_0$ (dBm)},
            ylabel = {Maximum Minimum Rate (bps/Hz)},
            ymin = 0,
            ymax = 0.04,
            xmin = 20,
            xmax = 40,
            grid = major,
            legend cell align = {left},
            legend style={at={(0,1)},anchor=north west},
        ]
            \addplot[
            restrict x to domain=20:40,
      unbounded coords=discard,
                black,
                mark = o,
                line width = 1pt,
                style = solid,
            ]
            table{Final2/search_vsP0_loss0.dat};
            \addlegendentry{\scriptsize Search-based WPPAN}


           \addplot[            restrict x to domain=20:40,
                red,
                mark = o,
                line width = 1pt,
                style = solid,
            ]
            table{Final2/greedy_vsP0_loss0.dat};
            \addlegendentry{\scriptsize Greedy WPPAN}

            \addplot[            restrict x to domain=20:40,
                green,
                mark = o,
                line width = 1pt,
                style = solid,
            ]
            table{Final2/naive_vsP0_loss0.dat};
            \addlegendentry{\scriptsize Naive WPPAN}

                        \addplot[            restrict x to domain=20:40,
                blue,
                mark = o,
                line width = 1pt,
                style = solid,
            ]
            table{Final2/bench_vsP0.dat};
            \addlegendentry{\scriptsize MISO (Benchmark)}


                \addplot[            restrict x to domain=20:40,
      unbounded coords=discard,
                red,
                mark = o,
                line width = 1pt,
                style = dotted,
                every mark/.append style={solid}
            ]
            table{Final2/greedy_vsP0_loss01.dat};

          \addplot[            restrict x to domain=20:40,
      unbounded coords=discard,
                red,
                mark = o,
                line width = 1pt,
                style = densely dashed,
                every mark/.append style={solid}
            ]
        table{Final2/greedy_vsP0_loss001.dat};
                        \addplot[            restrict x to domain=20:40,
                black,
                mark = o,
                line width = 1pt,
                style = dotted,
                every mark/.append style={solid}
            ]
            table{Final2/search_vsP0_loss01.dat};

                                    \addplot[            restrict x to domain=20:40,
                black,
                mark = o,
                line width = 1pt,
                style = densely dashed,
                every mark/.append style={solid}
            ]
            table{Final2/search_vsP0_loss001.dat};

    \addplot[            restrict x to domain=20:40,
                green,
                mark = o,
                line width = 1pt,
                style = dotted,
                every mark/.append style={solid}
            ]
            table{Final2/naive_vsP0_loss01.dat};

            \addplot[            restrict x to domain=20:40,
                green,
                mark = o,
                line width = 1pt,
                style = densely dashed,
                every mark/.append style={solid}
            ]
            table{Final2/naive_vsP0_loss001.dat};

           \node[draw, fill=white, inner sep=3pt, anchor=north west, font=\scriptsize] at (axis cs:20,0.022){%
        \begin{tabular}{lll}
        \tikz{\draw[solid, thick, black] (0.5cm,0) -- (1cm,0);}  $\kappa=0$ dB/m\\
          \tikz{\draw[dashed, thick, black] (0.5cm,0) -- (1cm,0);}  $\kappa=0.01$ dB/m\\
          \tikz{\draw[dotted, thick, black] (0.5cm,0) -- (1cm,0);}  $\kappa=0.1$ dB/m
        \end{tabular}
      };

        \end{axis}
    \end{tikzpicture}

%% file: vsUsers.tex
\begin{tikzpicture}
        \begin{axis}[
            scaled ticks=false, 
            tick label style={/pgf/number format/fixed},
            width=0.75\linewidth,
            xlabel = {Number of Users},
            ylabel = {Maximum Minimum Rate (bps/Hz)},
            ymin = 0,
            ymax = 0.14,
            xmin = 2,
            xmax = 10,
            xtick={1,2,3,4,5,6,7,8,9,10},
             ytick = {0,0.02,0.04,0.06,0.08, 0.1,0.12,0.14, 0.16, 0.20, 0.25},
            grid = major,
            legend cell align = {left},
            legend style={at={(1,1)},anchor=north east},
        ]

           \addplot[
                black,
                mark = o,
                line width = 1.5pt,
                style = solid,
            ]
            table{Final2/search_vsUsers_loss0.dat};
            \addlegendentry{\scriptsize Search-based WPPAN};


            \addplot[
                red,
                mark = o,
                line width = 1pt,
                style = solid,
            ]
            table{Final2/greedy_vsUsers_loss0.dat};
            \addlegendentry{\scriptsize Greedy WPPAN};


            \addplot[
                green,
                mark = o,
                line width = 1pt,
                style = solid,
            ]
            table{Final2/naive_vsUsers_loss0.dat};
            \addlegendentry{\scriptsize Naive WPPAN};
                    \addplot[
                blue,
                mark = o,
                line width = 1pt,
                style = solid,
                every mark/.append style={solid}
            ]
            table{Final2/bench_vsUsers.dat};
             \addlegendentry{\scriptsize MISO};

                        \addplot[
                blue,
                mark = o,
                line width = 1pt,
                style = densely dashed,
                every mark/.append style={solid}
            ]
            table{Final2/bench_vsUsers_n10_50reps.dat};


             \addplot[
                red,
                mark = o,
                line width = 1pt,
                style = densely dashed,
                every mark/.append style={solid}
            ]
            table{Final2/greedy_vsUsers_loss0_n10.dat};
            \addplot[
                green,
                mark = o,
                line width = 1pt,
                style = densely dashed,
                every mark/.append style={solid}
            ]
            table{Final2/naive_vsUsers_loss0_n10.dat};

                       \addplot[
                black,
                mark = o,
                line width = 1pt,
                style = densely dashed,
                every mark/.append style={solid}
            ]
            table{Final2/search_vsUsers_loss0_n10.dat};
            
\node[draw, fill=white, inner sep=3pt, anchor=north east, font=\scriptsize] at (axis cs:10,0.085){%
        \begin{tabular}{lll}
        \tikz{\draw[solid, thick, black] (0cm,0) -- (1cm,0);}  $N=4$ \\
          \tikz{\draw[dashed, thick, black] (0cm,0) -- (1cm,0);}  $N=10$ 
        \end{tabular}
      };         
        \end{axis}
    \end{tikzpicture}

%% file: histogram.tex
\begin{tikzpicture}
  \begin{axis}[
      ybar interval,
      bar width=1pt,
      width=0.7\linewidth,
      xlabel={Number of Activated Pinching Antennas},
      ylabel={Probability Density},
      ymin=0,
      ymax=0.35,
      xmin=0,
      xmax=10,
      grid =major,
      xtick={0,1,2,3,4,5,6,7,8,9,10},
      legend cell align = {left},
      legend style={at={(0,1)},anchor=north west}
  ]




        \addplot+[ybar interval,blue,
    fill=blue,mark=no,opacity=0.5] plot coordinates { (1, 0.00684064) (2, 0.0747) (3, 0.227768) (4, 0.3149) (5, 0.236382) (6, 0.105397) (7,0.028376) (8,0.005) (9, 0.000506714) (10, 0)};  
    \addlegendentry{\footnotesize Downlink};
    
    \addplot+[ybar interval,red,
    fill=red,mark=no,opacity=0.5] plot coordinates { (1, 0.0068) (2, 0.0488) (3, 0.1506) (4, 0.2762) (5, 0.2666) (6, 0.1682) (7,0.066) (8,0.0138) (9, 0.003) (10, 0)};
    \addlegendentry{\footnotesize Uplink}

    
    \end{axis}
\end{tikzpicture}

%% file: arxiv.bbl
\begin{thebibliography}{10}
\providecommand{\url}[1]{#1}
\csname url@samestyle\endcsname
\providecommand{\newblock}{\relax}
\providecommand{\bibinfo}[2]{#2}
\providecommand{\BIBentrySTDinterwordspacing}{\spaceskip=0pt\relax}
\providecommand{\BIBentryALTinterwordstretchfactor}{4}
\providecommand{\BIBentryALTinterwordspacing}{\spaceskip=\fontdimen2\font plus
\BIBentryALTinterwordstretchfactor\fontdimen3\font minus
  \fontdimen4\font\relax}
\providecommand{\BIBforeignlanguage}[2]{{%
\expandafter\ifx\csname l@#1\endcsname\relax
\typeout{** WARNING: IEEEtran.bst: No hyphenation pattern has been}%
\typeout{** loaded for the language `#1'. Using the pattern for}%
\typeout{** the default language instead.}%
\else
\language=\csname l@#1\endcsname
\fi
#2}}
\providecommand{\BIBdecl}{\relax}
\BIBdecl

\bibitem{psomas}
C.~Psomas, K.~Ntougias, N.~Shanin, D.~Xu, K.~Mayer, N.~M. Tran,
  L.~Cottatellucci, K.~W. Choi, D.~I. Kim, R.~Schober, and I.~Krikidis,
  ``{Wireless Information and Energy Transfer in the Era of 6G
  Communications},'' \emph{Proc. IEEE}, vol. 112, no.~7, pp. 764--804, 2024.

\bibitem{vranas}
C.~K. Vranas, P.~S. Bouzinis, V.~K. Papanikolaou, P.~D. Diamantoulakis, and
  G.~K. Karagiannidis, ``{On the Gain of NOMA in Wireless Powered Networks With
  Circuit Power Consumption},'' \emph{IEEE Commun. Lett.}, vol.~23, no.~9, pp.
  1657--1660, 2019.

\bibitem{ruizhang}
H.~Ju and R.~Zhang, ``{Throughput Maximization in Wireless Powered
  Communication Networks},'' \emph{IEEE Trans. Wireless Commun.}, vol.~13,
  no.~1, pp. 418--428, 2014.

\bibitem{elenaMIMO}
E.~Boshkovska, D.~W.~K. Ng, N.~Zlatanov, A.~Koelpin, and R.~Schober, ``{Robust
  Resource Allocation for MIMO Wireless Powered Communication Networks Based on
  a Non-Linear EH Model},'' \emph{IEEE Trans. Commun.}, vol.~65, no.~5, pp.
  1984--1999, 2017.

\bibitem{yu}
Y.~Zheng, S.~A. Tegos, Y.~Xiao, P.~D. Diamantoulakis, Z.~Ma, and G.~K.
  Karagiannidis, ``{Zero-Energy Device Networks With Wireless-Powered RISs},''
  \emph{IEEE Trans. Veh. Technol.}, vol.~72, no.~10, pp. 13\,655--13\,660,
  2023.

\bibitem{fluid}
\BIBentryALTinterwordspacing
X.~Lin, Y.~Zhao, H.~Yang, and J.~Hu, ``{Performance Analysis of Fluid Antenna
  Multiple Access Assisted Wireless Powered Communication Network},'' 2025.
  [Online]. Available: \url{https://arxiv.org/abs/2501.13405}
\BIBentrySTDinterwordspacing

\bibitem{ding2024flexible}
Z.~Ding, R.~Schober, and H.~Vincent~Poor, ``{Flexible-Antenna Systems: A
  Pinching-Antenna Perspective},'' \emph{IEEE Trans. Commun.}, pp. 1--1, 2025.

\bibitem{tegos2025pinching}
S.~A. Tegos, P.~D. Diamantoulakis, Z.~Ding, and G.~K. Karagiannidis, ``{Minimum
  Data Rate Maximization for Uplink Pinching-Antenna Systems},'' \emph{IEEE
  Wireless Commun. Lett.}, pp. 1--1, 2025.

\bibitem{zhou2025}
\BIBentryALTinterwordspacing
G.~Zhou, V.~Papanikolaou, Z.~Ding, and R.~Schober, ``{Channel Estimation for
  mmWave Pinching-Antenna Systems},'' 2025. [Online]. Available:
  \url{https://arxiv.org/abs/2504.09317}
\BIBentrySTDinterwordspacing

\bibitem{bozanis2025}
\BIBentryALTinterwordspacing
D.~Bozanis, V.~K. Papanikolaou, S.~A. Tegos, and G.~K. Karagiannidis,
  ``{Cram\'er-Rao Bounds for Integrated Sensing and Communications in
  Pinching-Antenna Systems},'' 2025. [Online]. Available:
  \url{https://arxiv.org/abs/2505.01333}
\BIBentrySTDinterwordspacing

\bibitem{khalili2025}
\BIBentryALTinterwordspacing
A.~Khalili, B.~Kaziu, V.~K. Papanikolaou, and R.~Schober, ``{Pinching
  Antenna-enabled ISAC Systems: Exploiting Look-Angle Dependence of RCS for
  Target Diversity},'' 2025. [Online]. Available:
  \url{https://arxiv.org/abs/2505.01777}
\BIBentrySTDinterwordspacing

\bibitem{docomo}
H.~O.~Y. Suzuki and K.~Kawai, ``{Pinching antenna: Using a dielectric waveguide
  as an antenna},'' \emph{NTT DOCOMO Technical Journal}, 2022.

\bibitem{mmWaveEH}
T.~X. Tran, W.~Wang, S.~Luo, and K.~C. Teh, ``{Nonlinear Energy Harvesting for
  Millimeter Wave Networks With Large-Scale Antennas},'' \emph{IEEE Trans. Veh.
  Technol.}, vol.~67, no.~10, pp. 9488--9498, 2018.

\end{thebibliography}
